\documentclass{PoS}

\title{MAGIC VHE Gamma-Ray Observations Of Binary Systems}

\ShortTitle{MAGIC VHE Gamma-Ray Observations Of Binary Systems}

\author{\speaker{D. Hadasch}\thanks{Corresponding author}\\
        Institute for Cosmic Ray Research, The University of Tokyo, 5-1-5 Kashiwanoha, Kashiwa-city, Chiba, 277-8582, Japan\\
        E-mail: \email{hadasch@icrr.u-tokyo.ac.jp}}

\author{W.~Bednarek$^{a}$, O.~Blanch$^{b}$, J.~Cortina$^{b}$, E.~de O\~na Wilhelmi$^{c}$, A.~Fern\'andez-Barral$^{b}$, R.~L\'opez-Coto$^{d}$, A.~L\'opez-Oramas$^{e}$, E.~Moretti$^{f}$, P.~Munar-Adrover$^{g}$, J.~M.~Paredes$^{h}$, M.~Rib\'o$^{h}$, D.~F.~Torres$^{i}$, J.~Sitarek$^{a}$ (for the MAGIC Collaboration) and J.~Casares$^{e}$\\
		$^{a}$ University of \L\'od\'z, PL-90236 Lodz, Poland\\
		$^{b}$ IFAE-BIST, Campus UAB, 08193 Bellaterra (Barcelona), Spain\\
		$^{c}$ CSIC/IEEC, E-08193 Barcelona, Spain\\
		$^{d}$ Max-Planck-Institut fur Kernphysik, 69029 Heidelberg, Germany\\
		$^{e}$ IAC and Universidad de La Laguna, Dpto. Astrof\'isica, E-38200/E-28206 La Laguna, Tenerife, Spain\\
		$^{f}$ Max-Planck-Institut f\"ur Physik, D-80805 M\"unchen, Germany\\
		$^{g}$ Unitat de F\'isica de les Radiacions, Departament de F\'isica, and CERES-IEEC, Universitat Aut\`onoma de Barcelona, E-08193 Bellaterra, Spain\\
		$^{h}$ Universitat de Barcelona, ICC, IEEC-UB, E-08028 Barcelona, Spain\\
		$^{i}$ ICREA and Institute for Space Sciences (CSIC/IEEC), E-08193 Barcelona, Spain \\}

\newcommand{\lsi}{LS\,I\,+61$^{\circ}$303}
\newcommand{\sip}{SS\,433}

\newcommand{\cyg}{V404\,Cygni}
\newcommand{\fermilat}{\textit{Fermi}-LAT}

\abstract{
There are several types of Galactic sources that can potentially accelerate charged particles up to GeV and TeV energies.
We present here the results of our observations of the source class of $\gamma$-ray binaries and the subclass of binary systems known as novae with the MAGIC telescopes.
Up to now novae were only detected in the GeV range. This emission can be interpreted in terms of an inverse Compton process of electrons accelerated in a shock. In this case it is expected that protons in the same conditions can be accelerated to much higher energies. Consequently they may produce a second component in the $\gamma$-ray spectrum at TeV energies.

The focus here lies on the four sources: nova V339 Del, \sip, \lsi\ and \cyg. The binary system \lsi\ was observed in a long-term monitoring campaign for 8 years. We show the newest results on our search for superorbital variability, also in context with contemporaneous optical observations. Furthermore, we present the observations of the only super-critical accretion system known in our galaxy: \sip.
Finally, the results of the follow-up observations of the microquasar V404 Cygni during a series of outbursts in the X-ray band and the ones of the nova V339 Del will be discussed in these proceedings.}

\FullConference{35th International Cosmic Ray Conference --- ICRC2017\\
		10--20 July, 2017\\
		Bexco, Busan, Korea}

\begin{document}

\section{INTRODUCTION}
The MAGIC experiment dedicates a significant fraction of observation time to binary systems in our Galaxy. Please find a detailed description of the experiment and its performance in \cite{aleksic2016_performance}.
The systems presented here are either $\gamma$-ray binaries: a small class among the family of X-ray binaries emitting the largest part of their non-thermal luminosity at high ($\geq$10\,MeV) $\gamma$-ray energies; or microquasars: X-ray (accreting) binaries consisting of a black hole (BH), that orbits a stellar companion; or novae: nuclear explosions on a white dwarf in a binary system.
 Here we present results of several observation campaigns on the following sources: nova V339 Del, \sip, \lsi\ and \cyg. 
All these sources are binary systems, but they have different characteristics. We try to catch unique emission patterns with different strategies like follow-up campaigns on triggers from other wavelengths, joint observation programs with other Cherenkov experiments or long-term monitoring. 


\begin{figure}[ht]
\begin{center}
 \includegraphics[width=0.39\textwidth]{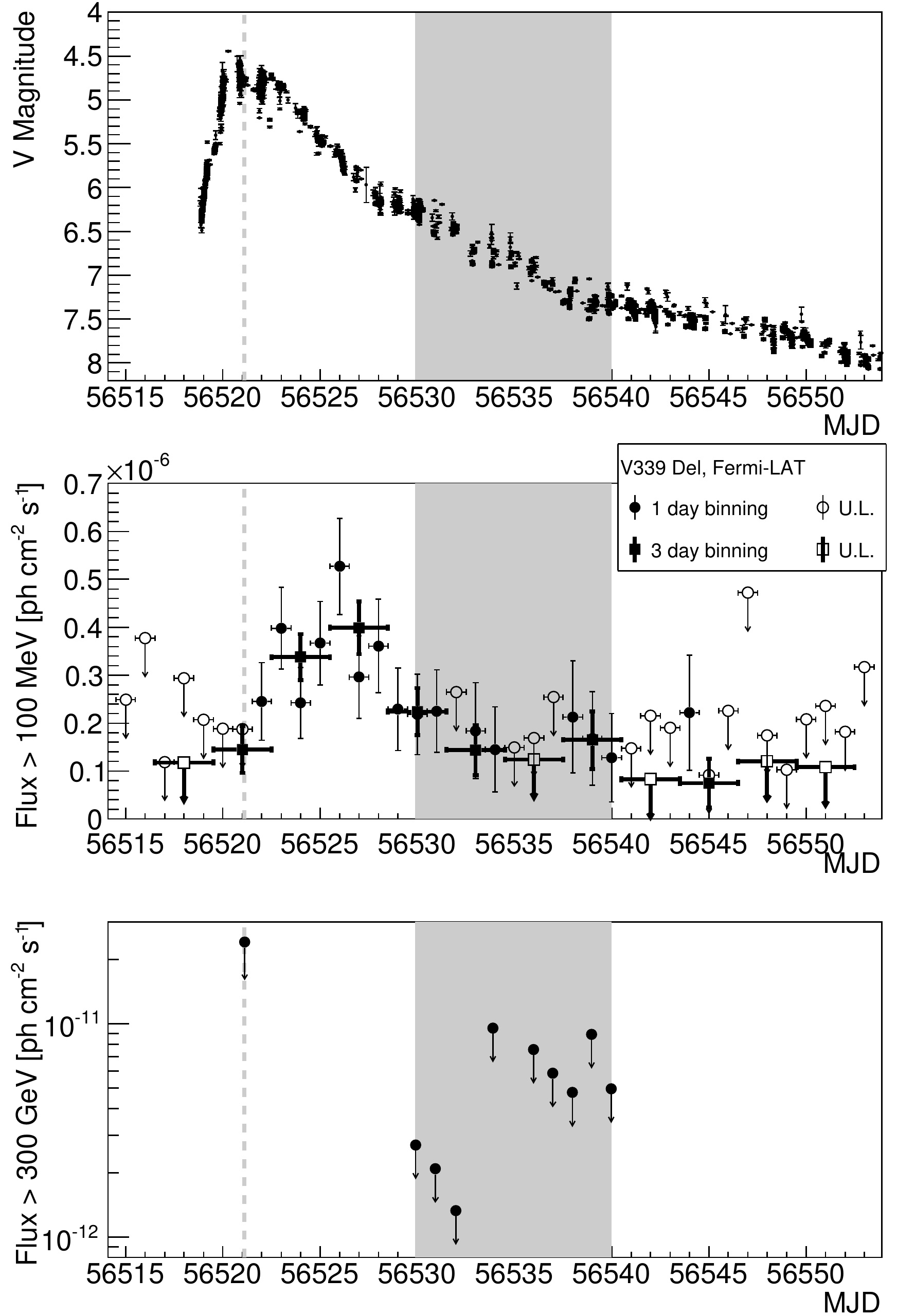}
 \includegraphics[width=0.49\textwidth]{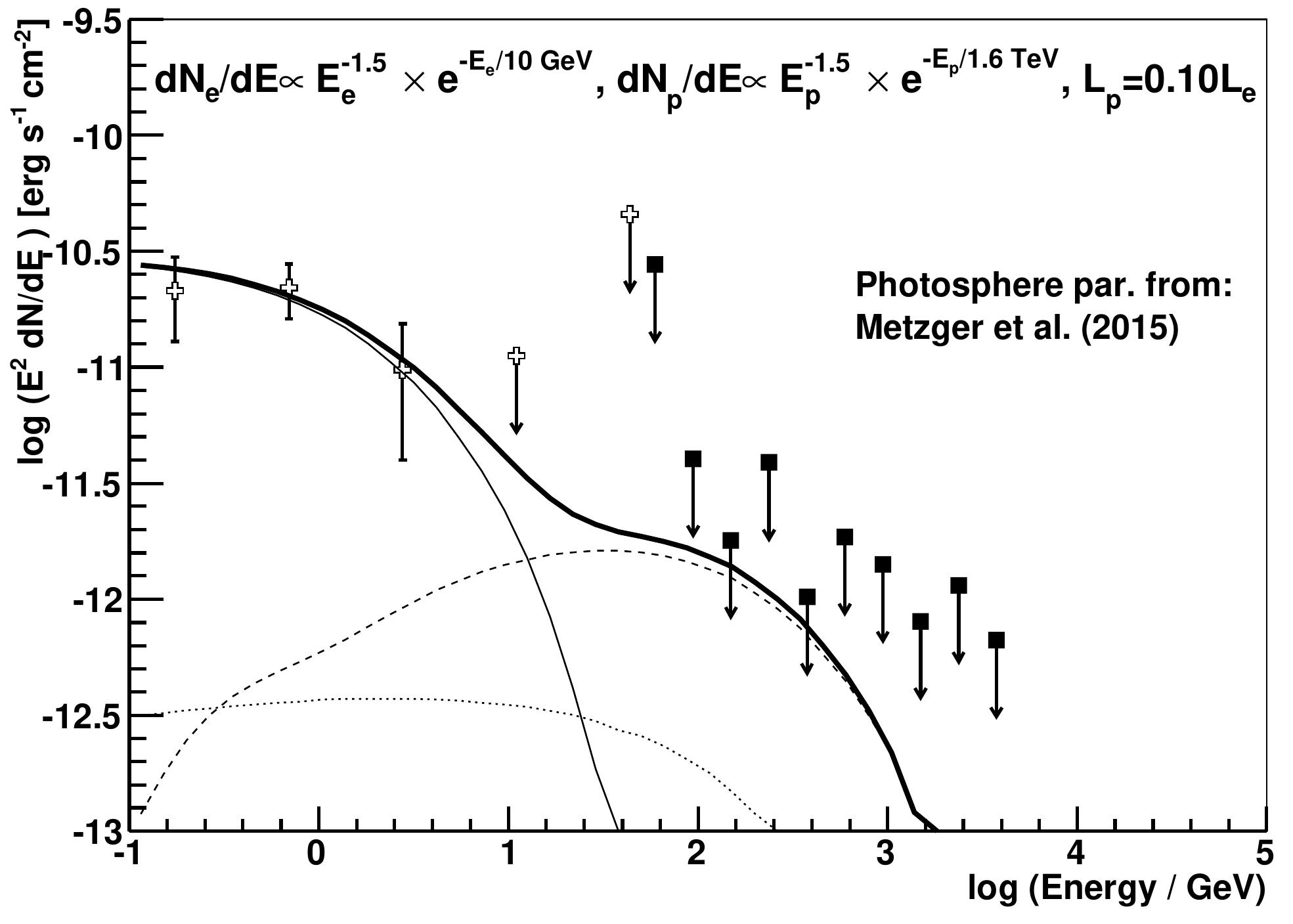}
  \caption{(a) Multiwavelength light curve of V339 Del during the outburst in August 2013.
Top panel: Optical observations in the V band obtained from the AAVSO-LCG service (http://www.aavso.org/lcg).
Middle panel: The \fermilat\ flux (filled symbols) and upper limits (empty symbols) above 100 MeV in 1-day (circles, thin lines) or 3-day (squares, thick lines) bins.
A 95\% C.L. flux upper limit is shown for time bins with TS$<$4.
Bottom panel: Upper limit on the flux above 300 GeV observed with MAGIC telescopes.
The gray band shows the observation nights with MAGIC.
The dashed gray line shows a MAGIC observation night affected by bad weather.
(b) Differential upper limits on the flux from V339 Del as measured by MAGIC (filled squares) and the flux measured by \fermilat\ (empty crosses) in the same time period, 2013 August 25 to September 4.
The thin solid line shows the IC scattering of thermal photons in the nova's photosphere.
The dashed line shows the $\gamma$-rays coming from the decay of $\pi^0$ from hadronic interactions of the relativistic protons with the nova ejecta.
The dotted line shows the contribution of $\gamma$-rays coming from IC of e$^+$e$^-$ originating from $\pi^+\pi^-$ decays.
Thick solid lines show the total predicted spectrum.
\label{fig:del_mwl}}
\end{center}
\end{figure}
\subsection{Nova V339 Del}
A classical nova is a thermonuclear runaway leading to the explosive ejection of the envelope accreted onto a white dwarf (WD) in a binary system in which the companion is either filling or nearly filling its Roche surface (see references in \cite{ahnen2015_novae}).
They are a type of cataclysmic variable, i.e. optically variable binary systems with mass transfer from a companion star to the WD.
In the last six years the Fermi Large Area Telescope (LAT) instrument detected GeV $\gamma$-ray emission from six novae (five classical novae and one symbiotic-like recurrent nova) \cite{cheung2016_novae}.
Since late 2012 the MAGIC collaboration has been conducting a nova follow-up program in order to detect a possible Very High Energy (VHE, $E>$100\,GeV) $\gamma$-ray component. 

Here we report on the observations performed with the MAGIC telescopes of V339 Del.
It was a fast, classical CO nova detected by optical observations on August 16 2013 (CBET \#3628), MJD 56520.
The nova was exceptionally bright reaching a magnitude of V$\sim 5\,$mag (see top panel of Fig.~\ref{fig:del_mwl}a), and it triggered follow-up observations at frequencies ranging from radio to VHE $\gamma$-rays.
No VHE $\gamma$-ray signal was found from the direction of V339 Del.
We computed night-by-night integral upper limits above 300\,GeV (see bottom panel of Fig.~\ref{fig:del_mwl}a) and differential upper limits for the whole good quality data set in bins of energy (Fig.~\ref{fig:del_mwl}b).
The $\gamma$-ray emission from V339 Del was first detected by \fermilat\ in a 1-day bin on August 18 2013 (see references in \cite{ahnen2015_novae}).
The emission peaked on August 22 2013 and entered a slow decay phase afterwards (Fig.~\ref{fig:del_mwl}a).
The GeV emission can be interpreted in terms of an inverse Compton process of electrons accelerated in a shock.
In this case it is expected that protons in the same conditions can be accelerated to much higher energies.
Consequently, they may produce a second component in the $\gamma$-ray spectrum at TeV energies.
We used the numerical code of \cite{sitarek2012} to model the \fermilat\ GeV spectrum and to compare the sub-TeV predictions with the MAGIC observations.
In Fig.~\ref{fig:del_mwl} we show the predictions for leptonic or hadronic spectra compared with the \fermilat\ and MAGIC measurements.
The \fermilat\ spectrum can be described mostly by IC scattering of the thermal photons in the nova's photosphere by electrons.
The expected hadronic component overpredicts the MAGIC observations at $\sim 100 \,$GeV by a factor of a few for the case of equal power of accelerated protons and electrons (i.e. $L_p=L_e$).
Using the upper limits from the MAGIC observations we can place the limit on $L_p\lesssim 0.15 L_e$.
Therefore, the total power of accelerated protons must be $\lesssim 15\%$ of the total power of accelerated electrons.
MAGIC will continue to observe promising $\gamma$-ray nova candidates in the following years.

\subsection{\sip}
\begin{figure}[ht]
\begin{center}
 \includegraphics[width=0.49\textwidth]{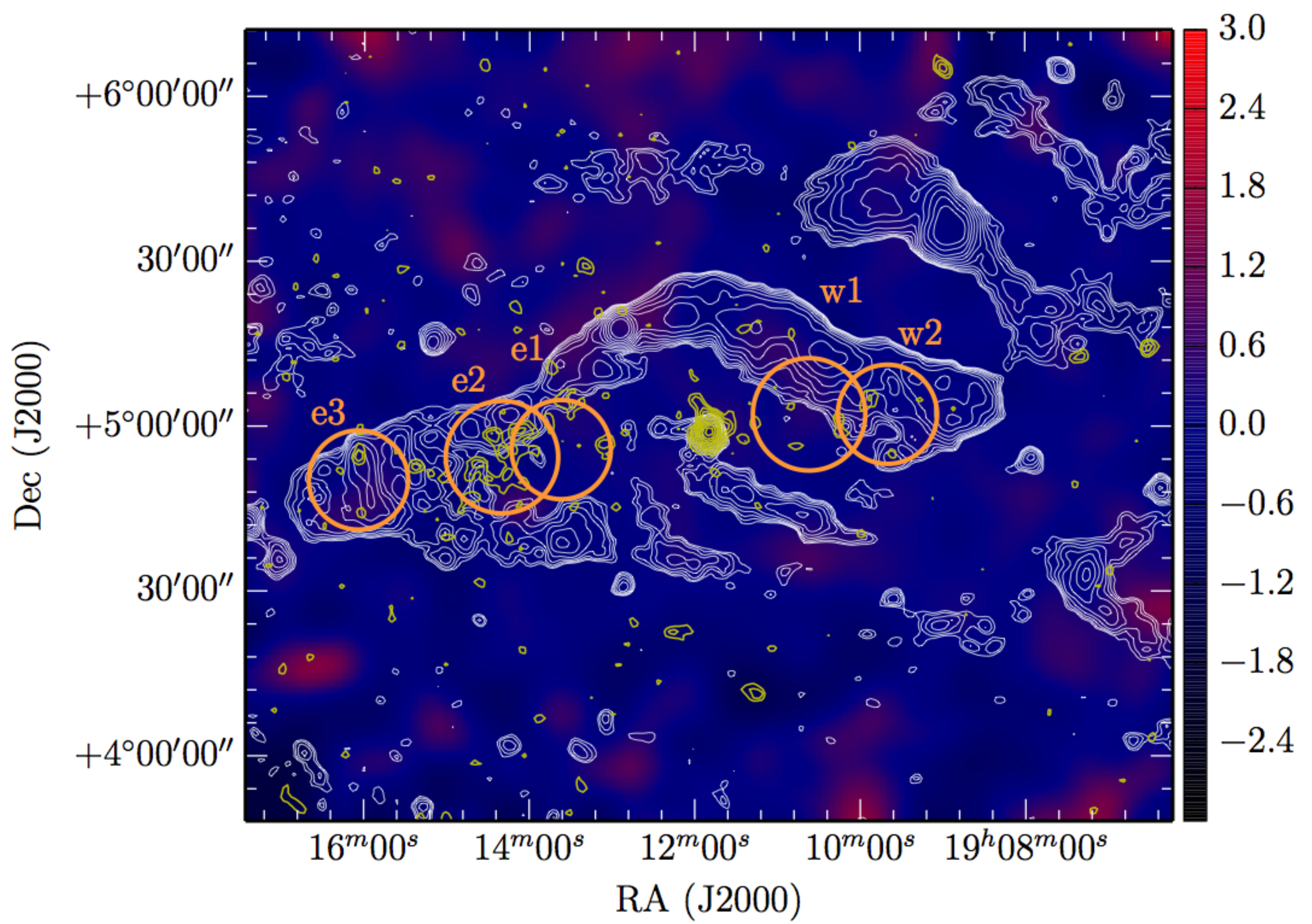}
  \caption{Skymap of the \sip\/W50 system as observed at E$\geq$250\,GeV by MAGIC. The color scale represents the excess events significance. GB6 4.85\,GHz radio contours (white \cite{1996ApJS..103..427G}) and ROSAT broadband X-ray contours (yellow \cite{1997ApJ...483..868S}) are over-plotted. Circles indicate the positions of interaction regions $w1, w2$ and $e1, e2, e3$.\label{fig:ss433}}
\end{center}
\end{figure}

\sip\ is an extremely bright microquasar with a bolometric luminosity of $L_{\rm bol} \sim {10}^{40}$ erg s$^{-1}$ and with the most powerful jets known in our Galaxy with $L_{\rm jet} \lesssim {10}^{39}$ erg s$^{-1}$ \cite{2002SSRv..102...23C,1998AJ....116.1842D}.
It is embedded in the W50 nebula (SNR G39.7--2.0 \cite{green2006}). Its present morphology is thought to be the result of the interaction between the jets of \sip\ and the surrounding medium. This scenario is supported by the position of SS433 at the center of W50, the elongation of the nebula in the east-west direction along the axis of precession of the jets and the presence of radio, IR, optical and X-ray emitting regions also aligned with the jet precession axis (see Fig.~\ref{fig:ss433}).
We explore the capability of \sip\ to emit VHE $\gamma$ rays during periods in which the expected flux attenuation due to periodic eclipses (P$_{\rm orb} \sim$ 13.1\,d) and precession of the circumstellar disk (P$_{\rm prec}\sim$ 162\,d) periodically covering the central binary system is expected to be at its minimum \cite{2008APh....28..565R}.
Our observations do not show any significant VHE emission neither for the central source \sip\ nor for any of the interaction regions with the W50 nebula $e1, e2, e3, w1$ and $w2$ (see Fig.~\ref{fig:ss433}).
Detailed results of the whole campaign together with the H.E.S.S. experiment will be published elsewhere soon.

\subsection{\lsi}
\lsi\ {(= V615 Cas )} is a $\gamma$-ray binary composed of a rapidly
rotating Be star of spectral type B0Ve with a circumstellar
disk and a compact object of unknown nature. The compact object, either a neutron star
(NS) or a stellar-mass black hole (BH), has an eccentric orbit ($e = 0.54 \pm 0.03$) with a period of
26.4960(28) days.

Orbit-to-orbit variability has been associated with a super-orbital modulation. This was first proposed based on centimeter radio variations that show approximately sinusoidal modulation over 1667 $\pm$ 8 days. A similar long-term behavior has recently been suggested {for X-rays} {(3 -- 30\,keV observed with \textit{RXTE}}), hard X-rays {(18 -- 60\,keV observed with \textit{INTEGRAL}}) 
and HE $\gamma$-rays {(100 MeV -- 300\,GeV observed with \textit{Fermi}/LAT}) (see references in \cite{2016A&A...591A..76A}). 
%


MAGIC observed \lsi\ as part of a multi-wavelength campaign between August 2010 and September 2014. All TeV data were
obtained with the MAGIC stereoscopic system, except for January 2012, when MAGIC-I
was inoperative. 
Data were taken during the orbital phase range $\phi$ = 0.5 -- 0.75  to scan the complete trend of the periodical
outburst peak of the TeV emission, with the aim of detecting a putative long-term modulation.
Contemporaneous observations with MAGIC and LIVERPOOL were performed during orbital phases 0.75 -- 1.0, which
are the phases where sporadic VHE emission had been detected and which does not seem to present yearly periodical variability of the flux level \cite{lsi_flux_states}. Since the fluxes in this orbital period are not influenced by the long-term modulation, changes in the relative optical and TeV 
fluxes are larger and easier to measure. The aim of these contemporaneous observations is to search for (anti-)correlation between the mass-loss rate of the Be star and the TeV emission.
 
\begin{figure}[ht]
\begin{center}
 \includegraphics[width=0.49\textwidth]{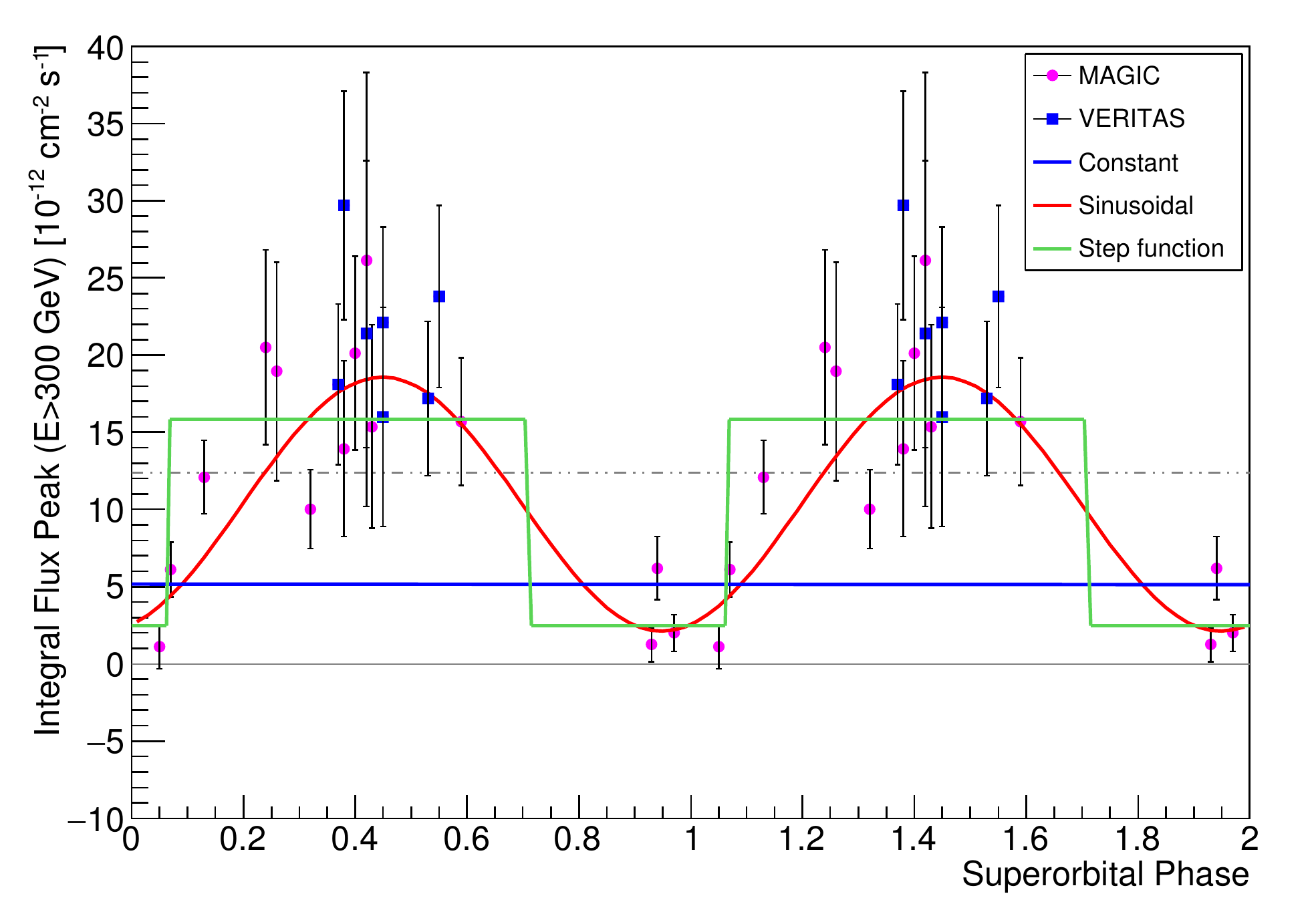}
 \includegraphics[width=0.39\textwidth]{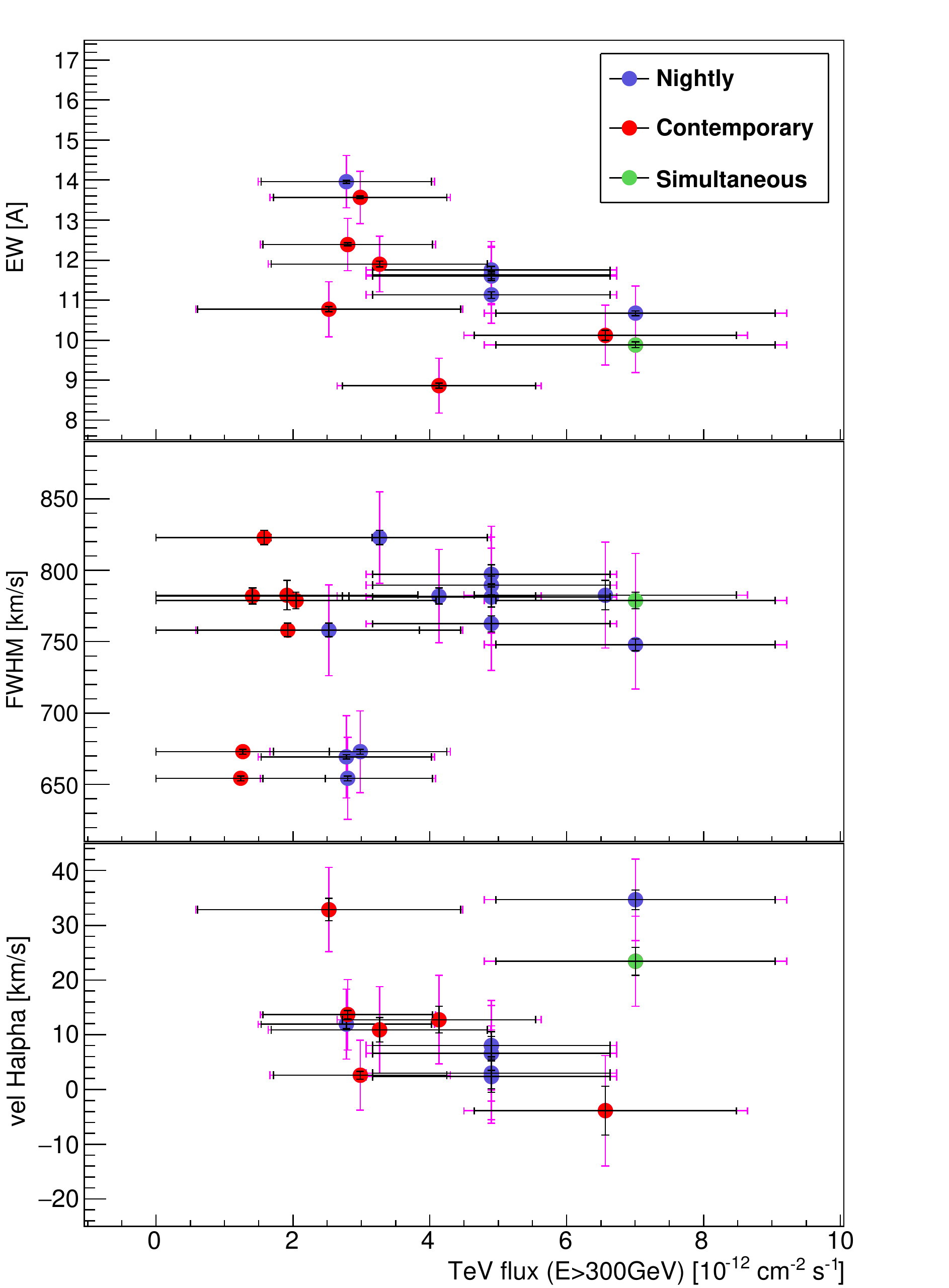}
\caption{Each data point represents the peak flux emitted in one orbital period during orbital phases 0.5 -- 0.75 and is folded into the super-orbit of 1667 days known from radio observations \cite{Gregory02}. MAGIC (magenta) and VERITAS (blue) points were fit with a sinusoidal (solid red line), with a step function (solid green line), and with a constant (solid blue line). The gray dashed line represents $10\%$ of the Crab Nebula flux, the gray solid line the zero level for reference.
(b) Correlations between the TeV flux obtained by MAGIC and the $H\alpha$ parameters (from top to bottom: EW, FWHM, and centroid velocity (vel)) measured by the  LIVERPOOL Robotic Telescope for the orbital interval 0.75 -- 1.0. Only TeV data points with a significance higher than 1$\sigma$ have been considered. Each data point represents a ten-minute observation in the optical and a variable integration in the TeV regime: nightly (blue), contemporary (red), and strictly simultaneous data (green). Black error bars represent statistical uncertainties, while systematic uncertainties are plotted as magenta error bars.
\label{fig:superorbit_MAGIC-VERITAS}}
\end{center}
\end{figure}

All archival data of \lsi\ recorded by MAGIC since its detection in 2006 \cite{2016A&A...591A..76A}
and the data from the observing campaigns presented were folded onto the superorbital period of 1667 days (Fig.~\ref{fig:superorbit_MAGIC-VERITAS}). The data were fit with a constant, a sinusoidal and to two-emission levels (step function). 
The probability for a constant flux is negligible, 4.5 $\times 10^{-12}$. Assuming a sinusoidal signal, the fit probability reaches 8\% ($\chi^{2}$/dof = 27.2/18). 
The fit to a step function resulted in a fit probability of 7\% ($\chi^{2}$/dof = 26.4/17).
We furthermore quantified the probability that the improvement found when fitting a sinusoid or a step function instead of a constant is produced by chance. To obtain this probability, we considered the likelihood ratio test. In both cases this chance probability is $<2.5\times10^{-10}$, which is low.
This shows that the observed intensity distribution can be described by a high and a low state and with a smoother transition. We conclude that there is a super-orbital signature in the TeV emission of \lsi\ and that it is compatible with the 4.5-year radio modulation seen in other frequencies.

The correlation between the TeV flux measured by MAGIC and the H$\alpha$ parameters measured with the telescope (equivalent width (EW), full width at half maximum (FWHM), profile centroid velocity) {were determined} including statistical and systematic uncertainties and the weighted \textit{\textup{Pearson correlation coefficient}} \cite{Pearsoncoeff1917}. 
No statistically significant correlation was found for the sample at orbital phases $\phi$ = 0.75 -- 1.0. A hint of a correlation is observed, but its significance is low. A stronger correlation might be blurred as a result of the fast variability of the optical parameters on short timescales compared to the long exposure times required by MAGIC, and as a result of the relatively large uncertainties and small number of data points used for this analysis. Figure~\ref{fig:superorbit_MAGIC-VERITAS}b shows the H$\alpha$ measurements plotted against the TeV flux.

The main conclusions from this multi-year analysis of \lsi\ observations are: 

(1) We achieved a first detection of super-orbital
variability in the TeV regime. {Using} the new VHE data and the MAGIC and
VERITAS archival data, we found that the super-orbital signature of \lsi\
is consistent with the 1667-day radio period within 8$\%$.

(2) There is no statistically significant intra-day correlation between H$\alpha$ line properties and TeV emission, nor is there an obvious trend connecting the two frequencies. 


\subsection{\cyg}
The microquasar V404 Cygni (V404 Cyg), located  at a parallax distance of 
2.39$\pm 0.14$ kpc \cite{2009ApJ...706L.230M}, is a binary system of an
accreting stellar-mass black hole from a companion star. 
  The black hole mass estimation ranges from about 8 to 15 M$_{\odot}$, 
while the companion star mass is $0.7^{+0.3}_{-0.2}$ M$_{\odot}$ \cite{Casares94, Khargharia10, Shahbaz94}. The system inclination 
angle is 67$^{\circ}$ $^{+3}_{-1}$ \cite{Shahbaz94, Khargharia10}
and the
system orbital period is 6.5 days \cite{Casares94}.
This low-mass X-ray binary (LMXB) showed at least four periods of outbursting activity: the one that led to its discovery in 1989 detected by the Ginga X-ray satellite
\cite{Ginga}, two previous ones in 1938 and 1956 observed in optical and later associated with V404 Cyg \cite{Richter}, and the latest in 2015.

In June 2015, the system underwent an exceptional 
flaring episode. From the 15th to the end of June the 
bursting activity was registered by several hard X-ray 
satellites, like \textit{Swift} and INTEGRAL \cite{SwiftGCN,INTEGRALAtel}.
It reached a flux about 40 times larger than the Crab Nebula one 
in the 20--40 keV energy band \cite{V404Rodriguez2015}.
Triggered by the INTEGRAL alerts, MAGIC observed V404 Cyg for several nights between June 18th and 27th 2015, collecting more than 10 hours.  

\begin{figure}[ht]
\begin{center}
 	\includegraphics[width=0.72\columnwidth]{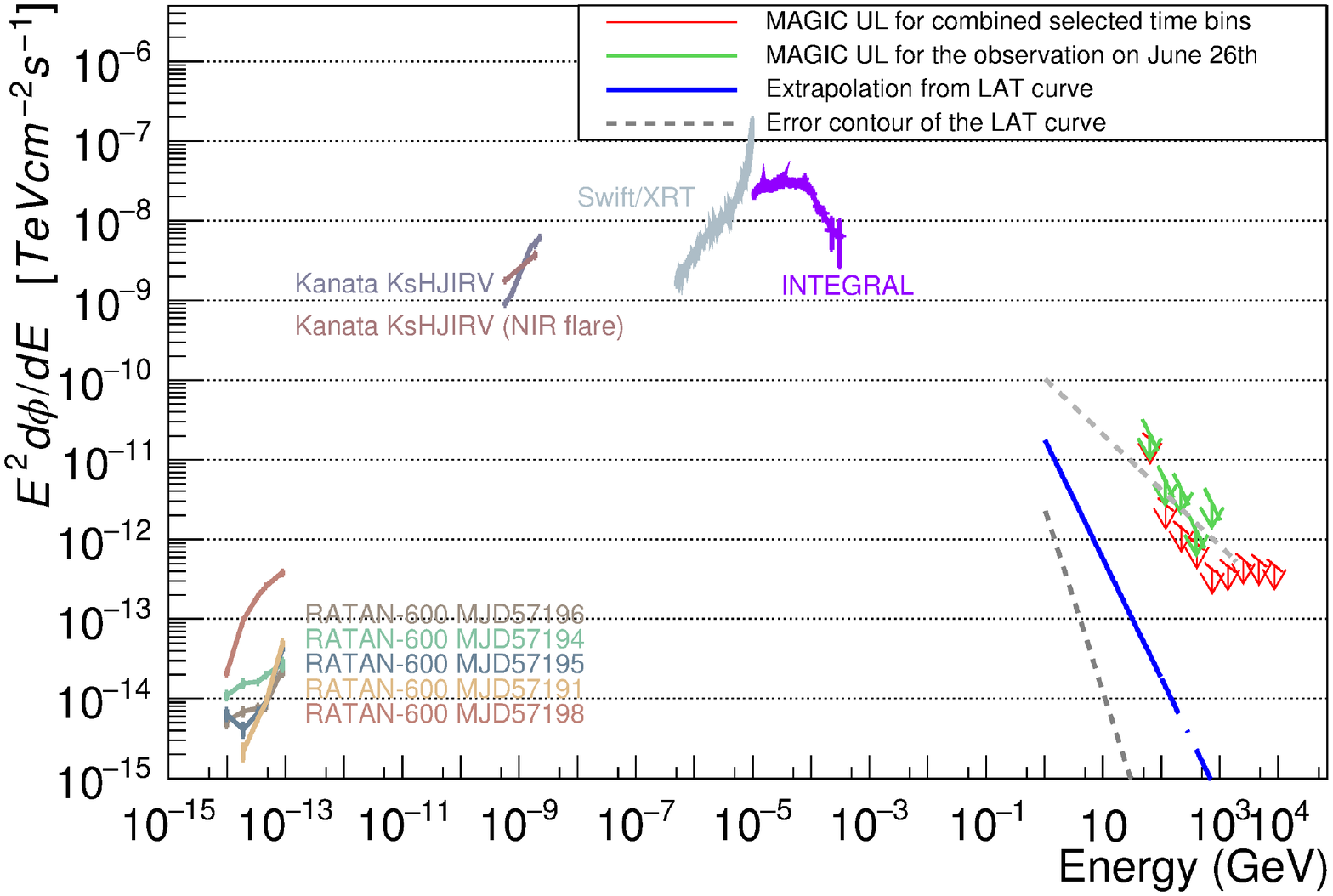}
     \caption{Multiwavelength spectral energy
     distribution of V404 Cyg during the June 
     2015 flaring period. In red, MAGIC ULs are
     given for the combined Bayesian block 
     time bins ($\sim$7 hours) for which a
     power-law function with photon index 
     2.6 was assumed. In green, MAGIC ULs for
     observations on June 26th, simultaneously
     taken with the {\it Fermi}-LAT hint
     \cite{V404_Fermi}. In this case, a photon
     index of 3.5 was applied following 
     {\it Fermi}-LAT results. All the MAGIC upper
     limits are calculated for a 95\% confidence
     level, considering also a 30\% systematic
     uncertainty. The extrapolation 
     of the {\it Fermi}-LAT spectrum is shown in
     blue with 1 $\sigma$ contour (gray dashed
     lines). In the X-ray regime, INTEGRAL (20-40\,keV) 
     \cite{V404Rodriguez2015} and 
     {\it Swift}-XRT (0.2-10\,keV)
     \cite{V404_Tanaka} data are
     depicted. At lower energies, 
     {\it Kanata}-HONIR optical and NIR data are
     shown, taken from \cite{V404_Tanaka}.}
     \label{fig:diff_UL}
\end{center}
\end{figure}

Most of the observations were performed
during the strongest hard X-ray flares. In total, MAGIC observed the microquasar for 8 
non-consecutive nights collecting more than 10 hours of data, some coinciding with observations at other energies.
To avoid an iterative search over different time bins, 
we assumed that the TeV flares were simultaneous to 
the X-ray ones. We defined the time intervals  
where we search for signal in the MAGIC data, 
to match those of the flares in the INTEGRAL 
light curve. We analysed the 
INTEGRAL-IBIS data (20--40 keV) publicly available 
with the \texttt{osa} software version 10.2\footnote{http://www.isdc.unige.ch/integral/analysis}. 
The time selection for the MAGIC analysis was performed running a Bayesian block \cite{2013ApJ...764..167S} analysis on the INTEGRAL light curve.
We searched for VHE gamma-ray emission stacking the MAGIC data of the selected time intervals ($\sim7$ hours).   
We found no significant emission in the $\sim7$  hour
sample, neither in any of the sub-samples considered. We then computed differential upper limits (see Figure \ref{fig:diff_UL}) 
for the observations assuming a power law spectral shape of index -2.6. 
The luminosity upper limits calculated for the full observation period,  
considering the source at a distance of 2.4 kpc, is $\sim 2 \times 10^{33}$ 
erg s$^{-1}$, in contrast with the extreme luminosity emitted in the X-ray band 
($\sim 2 \times 10^{38}$ erg s$^{-1}$, \cite{V404Rodriguez2015}) and other wavelengths.

Models predict TeV emission from this type of systems under efficient particle
acceleration on the jets \cite{1999MNRAS.302..253A, 2015ApJ_806_168Z} or strong hadronic jet component
\cite{VilaRomero}. If produced, VHE gamma rays may get extinct via pair creation in the vicinity of the emitting region. For gamma rays in an energy range 
between 200 GeV -- 1.25 TeV, the largest cross section
occurs with NIR photons.  
For a low-mass microquasar, like V404~Cyg, the contribution 
of the NIR photon field from the companion star (with a 
bolometric luminosity of $\sim10^{32}$ erg s$^{-1}$) is 
very low. During the period of flaring activity, disk and jet contributions are expected to dominate. During the outburst 
activity of June 2015, 
 the luminosity on the NIR regime can be estimated to $L_{NIR}=4.1\times 10^{34}$ erg s$^{-1}$. Assuming this luminosity, 
the gamma-ray 
opacity at a typical radius $r\sim 1\times 10^{10}$~cm may be relevant 
enough to avoid VHE emission above 200 GeV. Moreover, if IC on X-rays at 
the base of the jets ($r\lesssim 1\times 10^{10}$ cm) is produced, this 
could already prevent electrons to reach the TeV regime, unless the 
particle acceleration rate in V404\,Cyg is close to the maximum 
achievable including specific magnetic field conditions (see e.g. 
\cite{Khangulyan2008}). 
On the other hand, absorption becomes negligible for $r>1\times 10^{10}$ 
cm. Thus, if the VHE emission is produced in the same region as HE 
radiation ($r\gtrsim 1\times 10^{11}$ cm, to avoid X-ray absorption), 
then it would not be significantly affected by pair production
attenuation ($\sigma_{\gamma \gamma}< 1$). Therefore a VHE emitter at $r\gtrsim 1\times 10^{10}$ cm, along to the non-detection by MAGIC,
suggests either inefficient particle 
acceleration inside the V404\,Cyg jets or not enough energetics 
of the VHE emitter.

\section{ACKNOWLEDGMENTS}
We would like to thank the IAC for the excellent working conditions at the ORM in La Palma. We acknowledge the financial support of the German BMBF, DFG and MPG, the Italian INFN and INAF, the Swiss National Fund SNF, the European ERDF, the Spanish MINECO, the Japanese JSPS and MEXT, the Croatian CSF, and the Polish MNiSzW.

\end{document}